# Resonant Tunneling in Tri-layer 2H-MoTe$_2$ grown by Molecular Beam Epitaxy Coupled with layered WSe$_2$ carrier Reservoir

Abir Mukherjee[1,#], Kajal Sharma[1,#], Kamlesh Bhatt[2], Santanu Kandar[2], Rajendra Singh[2,3] and Samaresh Das[1,3,*]

[1] Centre for Applied Research in Electronics, Indian Institute of Technology Delhi, Delhi, 110016, India

[2] Dept. of Physics, Indian Institute of Technology Delhi, Delhi, 110016, India

[3] Dept. of Electrical Engineering, Indian Institute of Technology Delhi, Delhi, 110016, India

* Corresponding Author: samareshdas@care.iitd.ac.in

#: AM and KS contributed equally to this article

## Abstract

Here, we report a prominent quantum oscillation in the conductance of 2H-MoTe$_2$ based resonant tunneling structure. In this work, a n-WSe$_2$/HfO$_2$/i-MoTe$_2$/HfO$_2$/Au resonant tunneling device (RTD) with a symmetric & asymmetric double barrier has been fabricated using Molecular Beam Epitaxy (MBE) grown 2H-MoTe$_2$ and Chemical Vapor Deposition (CVD) grown 2H-WSe$_2$ along with theoretical modeling by adopting non-equilibrium green's function (NEGF) formalism. The impact of MoTe$_2$-quantum well widths equal, and above its excitonic Bohr radius (EBR ~ 0.7 nm) on resonant tunneling current is investigated at cryogenic temperatures. Such peak values increase with downscaling of the well width up to a certain value and then it decreases with further miniaturization. The corresponding maximum peak-to-valley current ratio (PVR) is estimated to be ~ 4 at 4K in the low voltage range for the very first time in MoTe$_2$ based RTD. Therefore, the present work may provide the route for fabrication of WSe$_2$/MoTe$_2$-based high performance resonant tunneling devices integrable with HEMT device for modern Qubit architecture operational at ultra-low temperatures.

# I. Introduction

Resonant tunneling phenomenon in semiconducting device structures has been exploited for multifaceted applications including quantum cascade lasers, Terahertz (THz) emitters & imaging, high speed memory, and digital circuits, as well as for Qubit generation in quantum computing and random number generation in recent times[1–5]. Such fascinating applications of the phenomenon have been possible through their implementation in novel nanoscale electronic & optoelectronic devices including the resonant tunneling diode, resonant tunneling bipolar transistors and FETs, TFETs, gate tunable quantum occupancy effect in junction less FET and double quantum dots[6–8]. Several approaches have been attempted to theoretically model such an interesting physical phenomenon, with varying degrees of validity level and accuracy. Such methods include Easki-Tsu formula, density matrix and Wigner function formulations, and Non-Equilibrium Green's function (NEGF) formalism[9–12]. Among such approaches, direct calculation of transmission coefficients is conceptually simple, however, it is a tedious process to calculate or compute and based on the assumption of only pure states to be present in the device. This contradicts the practical carrier transport in resonant tunneling which is a non-equilibrium process and therefore may have mixed states. In this context, Green's function approach can model the non-equilibrium transport of charges in active quantum device coupled to reservoirs, where all the statistical and phase breaking processes are incorporated through corresponding self-energy matrices[13]. Therefore, the NEGF approach has emerged as the most promising method for modeling quantum transport in electronic and optoelectronic devices[14]. Several novel physical phenomena including the negative differential resistance/ transconductance, ambipolar transport and resonant tunneling have been utilized to challenge the limitations of conventional FET devices. However, the appropriate design of such devices and their fabrication compatibility with the mainstream CMOS processing is yet to be achieved. In this context, the utilization of multi-resonant tunneling transport through controlled energy states in quantum wells/dots (QW/QD) can be an alternative approach to achieve multiple current threshold devices[12,15]. However, the transport properties of QW/QDs are very susceptible to their minute variation in morphology and composition, and therefore, the fabrication of reliable and reproducible quantum structures (QS) poses significant challenges. Also, the integration of such QSs in conventional CMOS process flow is cumbersome. In this context, the use of voltage-tunable quantum well/wire/dots where the eigenstates can be controlled by external biases such as electric field, magnetic field and/or electromagnetic field (e.g.

photon sources from THz to deep UV range) would be immensely beneficial[9,16–18]. Nowadays, 2D-materials (e.g. TMDCs) compatible Resonant tunneling device (RTD) devices are in huge demand[19–21]. Therefore, growth of high-quality TMDCs' ultra-thin films is obligatory for fabrication of such devices. In this present work, bulk-WSe$_2$, grown via Chemical Vapor Deposition and Tri-layer MoTe$_2$, grown via Molecular Beam Epitaxy (thickness ~ 2.5 nm) have been used as source reservoir and resonant tunneling medium respectively with integration of e-beam deposited HfO$_2$ as the tunnel junctions.

## II. Materials Growth and Characterization

To achieve 2D-TMDC materials compatible RTD architecture, chemical vapor deposition (CVD) and molecular beam epitaxy (MBE) have been utilized for the growth of WSe$_2$ and MoTe$_2$ respectively. N-type WSe$_2$ (bulk in nature ≥ 20 nm) had been grown via selenization of thinner W-film (≤ 10 nm) on SiO$_2$/Si substrate at optimized growth parameters such as temperature (650°C) and Ar+H$_2$ gas pressure (50 sccm) in the CVD system[22]. Raman spectroscopy, valence band spectra using XPS setup and HR-TEM measurements have been performed as shown in **Figure.1(a-d)** to confirm the material's growth with Raman shift 251 cm$^{-1}$ (E$^1_{2g}$), n-type doping characteristics determined from the energy gap of 0.935 eV between fermi level and valence band ($\Delta E_F^V$), and to identify (100) crystalline plane with 0.28 nm d-spacing respectively. To deposit the first insulation layer of 10 nm HfO$_2$ (being used tunnel barrier), electron beam evaporation has been used with emission current 12 mA under vacuum pressure 1.8 x 10$^{-7}$ mbar as shown in the AFM image (**Figure.2 (c, d)**). Thereafter, this sample was loaded in the process chamber of the MBE-system for the growth of Tri-layer MoTe$_2$. In this context, growth of MoTe$_2$ over HfO$_2$/WSe$_2$ was needed to optimize since growth on other substrates (such as Sapphire, SiO$_2$/Si etc.) are previously done by our group[23–25]. To avoid damage on prior films deposited, pre-annealing at the storage chamber was done at 150°C with ramp rate 12°C/min for 3hrs under UHV condition (~ 4 x 10$^{-10}$ mbar) to eliminate water molecules only. At the growth/process chamber under initial vacuum (~ 2 x 10$^{-11}$ mbar), to initiate the deposition, Te-cell was heated up to 1000°C (Top, with ramp rate 10°C/min) and 325°C (Bottom, with ramp rate 5°C/min). For deposition to be occurred, substrate was initially heated to 470°C (with ramp rate 10°C/min) and kept stable 450°C for further growth. To evaporate the Mo, e-beam emission current was raised to 75 mA (with 26.7A filament current at 10 kV supply voltage). Initially, Te-shutter was turned for some time to create the Te-reach environment. Once it happened, the e-beam shutter was turned open for the growth of MoTe$_2$. To verify the growth as well as to look after the reciprocal lattice since our intension to grown 2H-MoTe$_2$ phase, reflection high energy electron diffraction (RHEED) was monitored at supply voltage 12 kV and current 1.46 A.

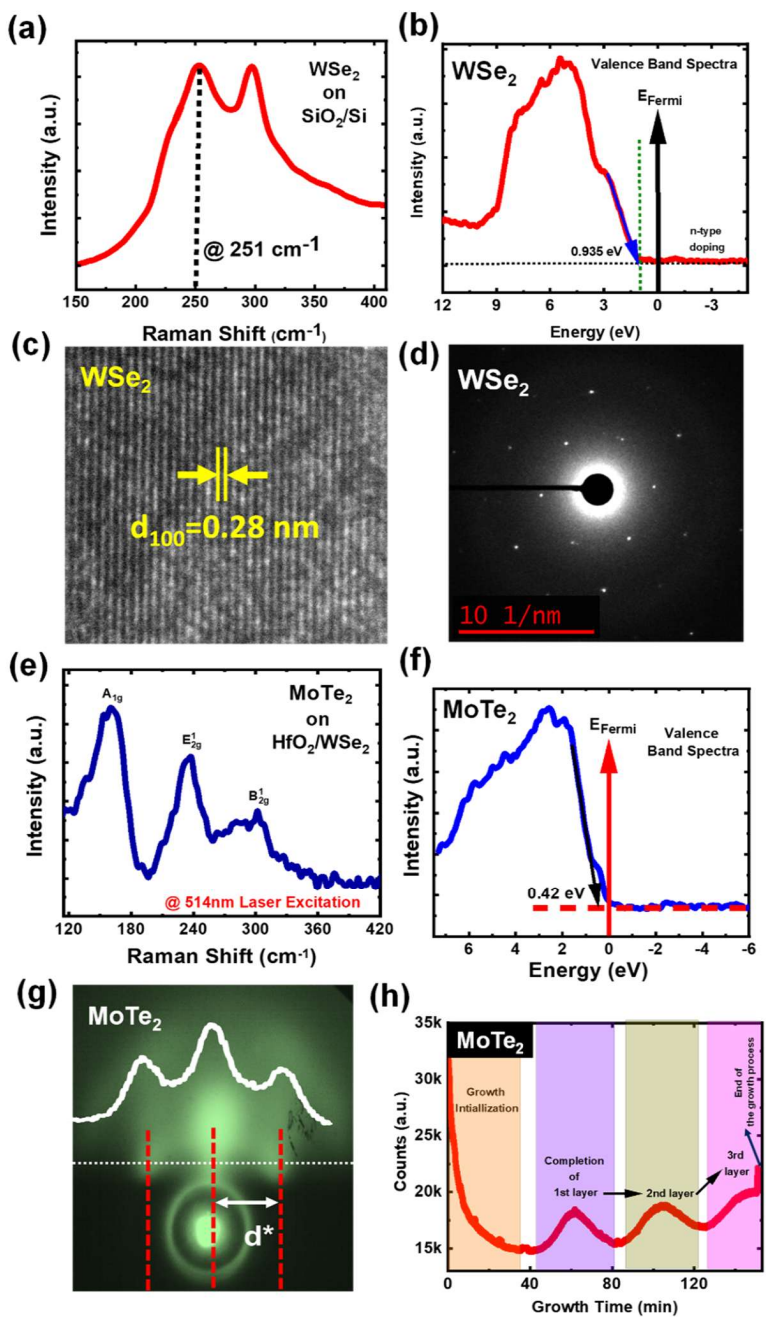

**Figure.1 (a) Raman spectroscopy at 514 nm laser excitation with five accumulations, (b) valence band spectra using the XPS set up, (c, d) HR-TEM image and SAED pattern using HR-TEM set up of the**

grown WSe$_2$ film via selenization of thin W-film, (e) Raman spectroscopy, (f) valence band spectra of MBE grown MoTe$_2$ and (g) RHEED pattern after the completion of the growth (80mins) and (h)RHEED oscillation confirming Tri-layer growth

From **Figure.1 (f)**, different in and out of plane vibrational modes (such A$_{1g}$, E$^1_{2g}$ and B$^1_{2g}$) are clearly observable at 168 cm$^{-1}$, 236 cm$^{-1}$ and 280 cm$^{-1}$ from the Raman spectroscopy at minimized laser power 10 mW with 5 accumulations, of grown 3L-MoTe$_2$ where B$^1_{2g}$ is sensitive to layer numbers and lattice strain[26]. Valence band spectra measurement confirms the intrinsic characteristics ($\Delta E^V_F$ = 0.42 eV considering the band gap ~ 0.9 eV) of the grown MoTe$_2$ film as shown in **Figure.1 (g)**. Film surface was monitored for the entire growth period by in-situ RHEED characterization. The cloudy diffused pattern shows the amorphous nature of prior deposited HfO$_2$ while transition in the RHEED pattern (from cloudy pattern to streak, **Figure.1 (g)**) took place when crystallinity arrived with layer-by-layer growth with growth rate 27 mins/1L (**Figure.1 (h)**). This streak directly provides the inherent insight of reciprocal lattice of MoTe$_2$.

### III. Device Fabrication and Measurement

Before processing the device fabrication, it was essential to measure the thickness of the deposited MoTe$_2$ (~ 2.5 nm from **Figure.2 (e, f)**) with confirmation of layer numbers arriving from cross-sectional TEM and RHEED oscillation (as shown above in **Figure. 1(h)**) as well.

To create double barrier resonant tunneling structure (**Figure.2 (a)**), 10 nm HfO$_2$ over MoTe$_2$ as depicted by AFM image, **Figure.2 (g, h)** was deposited via e-beam evaporation (with 12 mA emission current) under very high vacuum condition (1.4 x10$^{-7}$ mbar). Now to finalize the RTD fabrication, the top metal contact (Ti: Au:20nm:60nm) over WSe$_2$ other over top-HfO$_2$ were needed to be deposited. Therefore, laser writer (LW) was used to open a window using positive photo-resist, S1813 with intensity, 706 μJ/cm$^2$ with 405 nm laser wavelength and Reactive Ion Etching (RIE) was performed to etch down to WSe$_2$ using 15 sccm SF$_6$, 10 sccm O$_2$ and 5 sccm Ar at reduced RF power 40 W for 400 secs[22].

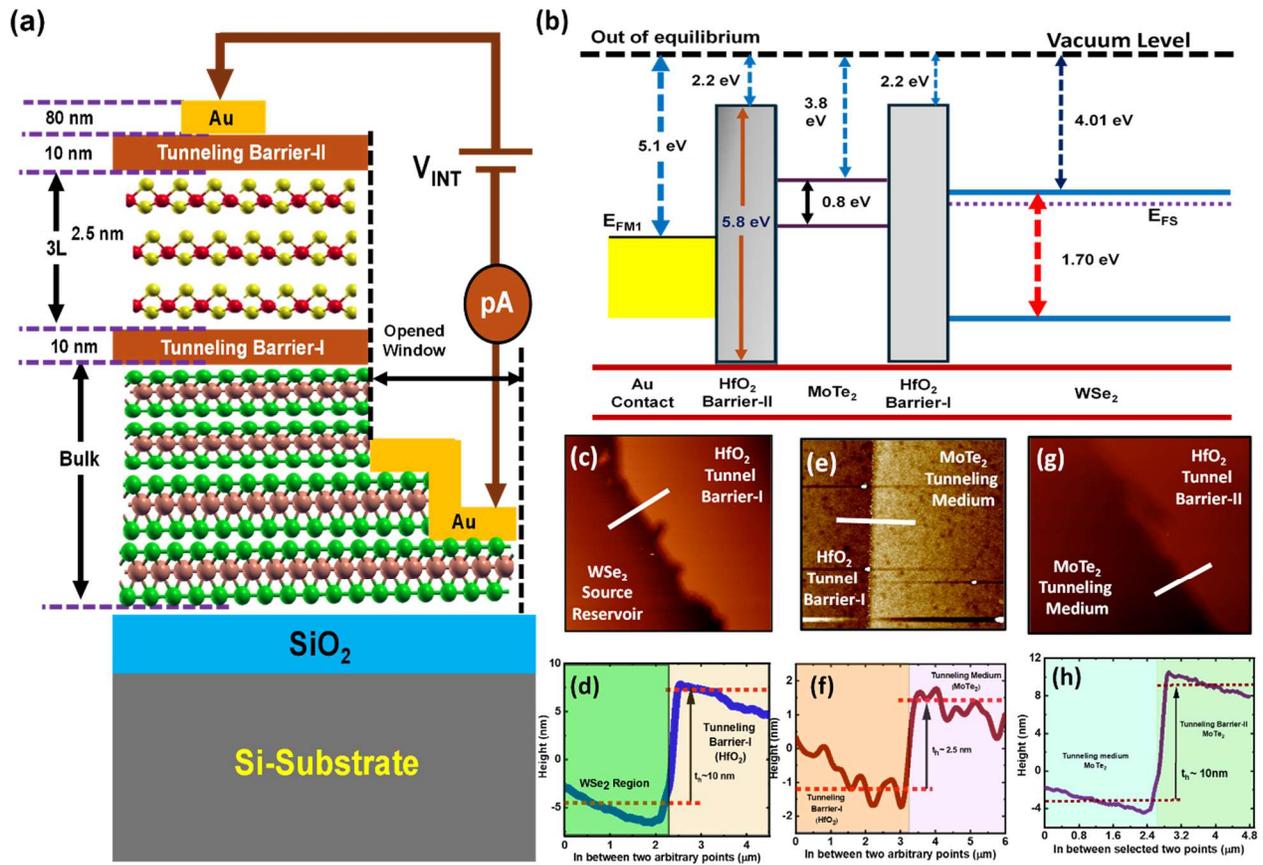

**Figure.2 (a)** Device Schematic, **(b)** band diagram of the device at out of equilibrium condition, AFM image & height estimation of **(c, d)** deposited 1st $HfO_2$ over $WSe_2$, **(e, f)** Grown $MoTe_2$ over $HfO_2$ and **(g, h)** 2nd $HfO_2$ over $MoTe_2$

Finally, using a shadow mask Ti: Au was deposited as schematically depicted in **Figure.2 (a)** confirmed with the optical & FESEM microscope images.

In device point of view, exhibition of $MoTe_2$ based resonant tunneling characteristics in lower voltage range is very hard to achieve because of its relatively high effective mass (i.e. less mobility), being a significant parameter to establish the strong phase coherence, comparable to other conventional TMDC materials. Demonstration of such resonant tunneling phenomenon in lower voltage range is very obligatory to achieve for precise quantum manipulations.

To perform the out of plane (OFP) electrical characterization, it was very essential to conduct the transport measurement at cryogenic temperatures (< 100K) since $MoTe_2$ shows relatively high mass density thereby providing a phonon dominated medium as discussed below in the theoretical background section.

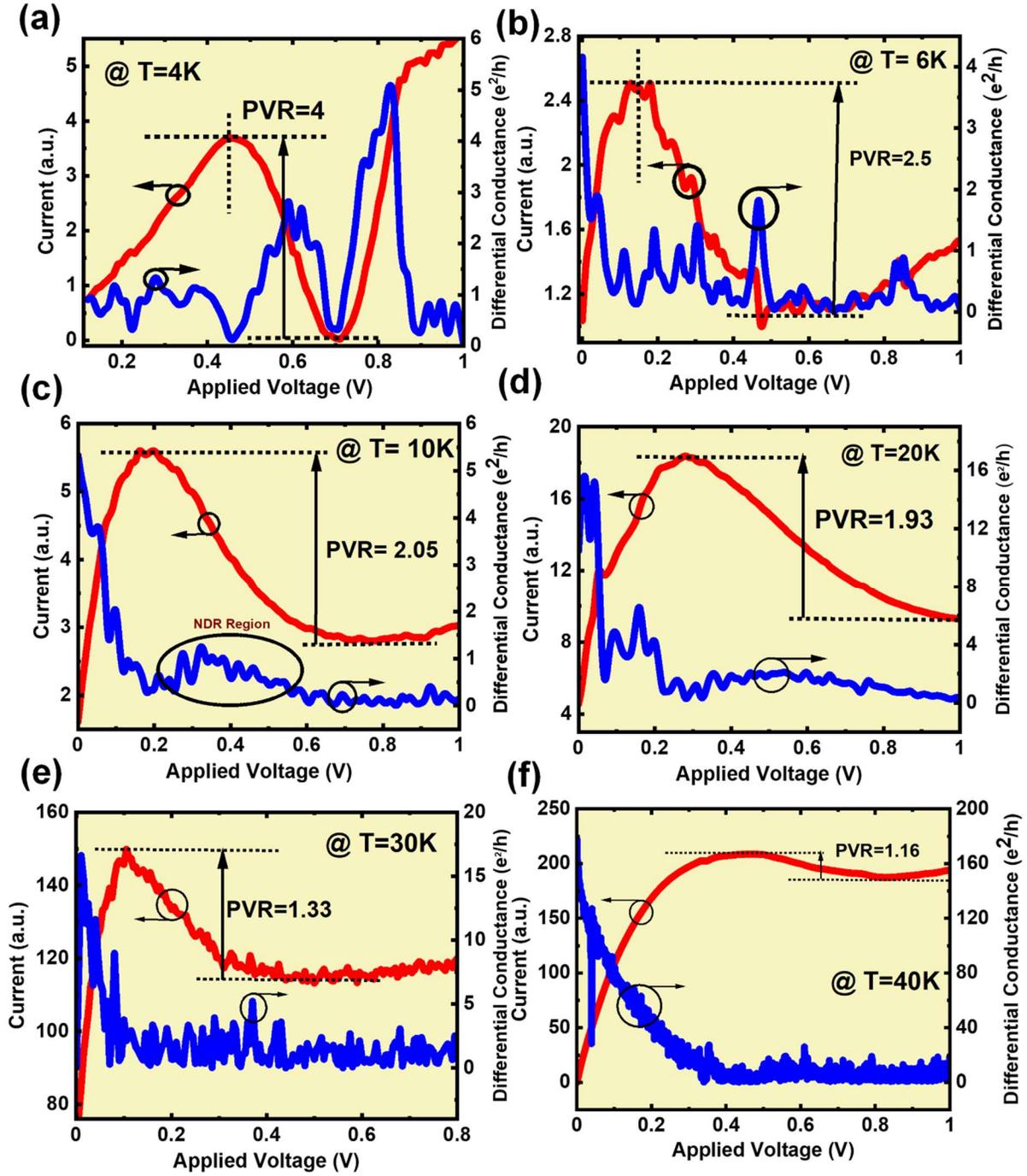

**Figure.3 Out of plane transport Characteristics: Current and differential conductance in linear scale at (a) 4K, (b) 6K, (c) 10K, (d) 20K, (e) 30K and (f) 40 K**

The resonant tunneling nature was identified by the negative (differential) resistance (NR or NDR) region as shown in **Figure.3 (a-f).** From this figure, it is clearly observable that NR/NDR region increases via increment in peak to valley ration as temperature approaches towards 4K. To unroll such NDR characteristics, -1 V to 1V was swept via quiet measurement technique. Within the temperature range 60 K to 300 K, the OFP transport characteristics were found to be linear, but as temperature got reduced beyond 60K, a negative resistance region was found to be appearing with gradual increment in peak to valley ration (PVR) where highest and lowest PVR, 4 and 1.16 occurred at 4 K and 40 K respectively. Plots for differential conductance are included with absolute format thereby NDR regions have been depicted by the positive peaks.

## IV. Quantum Field Theory Model & Result Discussion

To understand the fundamental aspects regarding the transport via out of plane quantized states of TMDC materials in the double barrier structure, it is essential to formulate the Hamiltonian with electron-phonon scattering which becomes largely significant due to high mass density in MoTe$_2$ (the tunneling medium). The second quantized Hamiltonian with modifications due to interaction arising from contact reservoirs, electron-electron scattering, electron-phonon scattering (fermion-boson interaction) can be written as [12,27–30],

$$H = \sum_i H^C_{ISO} c_i^+ c_i + \sum_{i,r} \left( \tau_{ir} c_i^+ c_r + \tau_{ri}^* c_r^+ c_i \right) + \sum_{i,j,\alpha} \left( \tilde{\tau}_{ij}^\alpha c_i^+ c_j b_\alpha + \tilde{\tau}_{ji}^{\alpha *} c_i^+ c_j b_\alpha^+ \right) \quad (1)$$

Where, $c_{i,r}$ and $b_\alpha$ are the fermionic field operator for electrons in the $i$ (active device, AD) and $r$ (top/bottom reservoirs) modes, and the bosonic field operator for phonons in $\alpha$ mode with angular frequency, $\omega_\alpha$ respectively. $\tau_{ir}$ represents the interaction strength between active device and reservoirs, while $\tilde{\tau}_{ij}^\alpha$ depicts how system (i.e. AD) interacts back with itself by means of electron-phonon coupling. The equ.(1) gives rise to equations of motion for the electrons and absorption/ emission of phonons.

The possible supply of electrons in a definite state 'i' in the conduction band is firstly from the state itself, secondly by absorption or emission of phonons, and thirdly from the reservoirs.

$$i\hbar \frac{d}{dt} c_i = H^C_{ISO} c_i + \sum_{j,\alpha} \left( \tilde{\tau}^\alpha_{ij} v_j b_\beta + \tilde{\tau}^\alpha_{ij} {}^* v_j b_\beta^+ \right) + \sum_r \tau_{ir} c_r \qquad (2)$$

The absorptions/emissions of phonons in a definite mode are added up with those associated with annihilation of electrons from a definite state and subsequent creation of another. The situation is expressed as,

$$i\hbar \frac{d}{dt} b_\alpha = E_\alpha b_\alpha + \sum_{i,j} \left( \tilde{\tau}^\alpha_{ij} {}^* c_i^+ c_j + \tilde{\tau}^\alpha_{ji} {}^* c_j c_i^+ \right) \qquad (3)$$

The phonon interaction strength can be written as follows:

$$\tilde{\tau} = D \Xi = D \left( \vec{\nabla} \cdot \vec{u} \right) \qquad (4)$$

Here, D is the deformation potential of the confining 2D-material (MoTe$_2$)[31] and $\Xi$ is the lattice strain due to phononic vibrations, derived from the lattice displacement of the oscillating atoms with wave vector $\vec{\upsilon}(\omega)$. Thus, the expression for the lattice displacement of the oscillating atoms is written below[32],

$$\vec{u} = \vec{u}_0 e^{i(\vec{\upsilon} \cdot \vec{r} - \omega t)} + \vec{u}_0^* e^{-i(\vec{\upsilon} \cdot \vec{r} - \omega t)} \qquad (5)$$

where, $\vec{u}_0$ can be calculated with approximation conserving Einstein-Debye mass oscillator as[33,34],

$$\vec{u}_0 = \hat{\vartheta} \sqrt{\frac{\hbar \omega f_{BE}(\omega, T)}{4 \rho \Omega \omega}} \begin{pmatrix} 1 \\ \pm 1 \end{pmatrix} \qquad (6)$$

Here, $\hat{\vartheta}$, $\Omega$, $\rho$ and $f_{BE}(\omega, T)$ are the polarization vector, normalization volume, mass density and Bose-Einstein distribution function for phonons at temperature, T. The "+" and "−" signs represent acoustic and optical phonons, respectively. Derived equations of motion for electrons have been solved in coupled mode space from the local non-equilibrium Green's function (as a function of energy) in the double barrier resonant tunneling medium coupled to different surroundings (as mentioned above) as given by[27–29,35],

$$[G^M(E)] = \left\{ [EI] - [H_{ISO}] - [\varepsilon^c_{k,y} + \varepsilon^c_{l,x}]\delta_{k,k';l,l'} - [\Sigma_{TR}] - [\Sigma_{BR}] - [\Sigma_{Scat}] \right\}^{-1} \quad (7)$$

$\varepsilon^c_{k,y}$ and $\varepsilon^c_{l,x}$ are the energy eigen values approaching from the lateral directions along which tunneling material offers degree of freedom. $\Sigma_{TR,BR}$ are the self-energies for top and bottom reservoir respectively. The local density of states (LDOS) occupied by electrons are therefore obtained to be[35,36],

$$[n(E)] = [G^n(E)] = \left(\frac{2}{2\pi a}\right)[G^M(E)]\left[\Sigma^{In}_{TR}(E) + \Sigma^{In}_{BR}(E) + \Sigma^{In}_{Scat}(E)\right][G^M(E)]^\dagger \quad (8)$$

where, $\quad \Sigma^{In}_{TR/BR}(E) = [\tau]^\dagger [n_{TR/BR}(E)][\tau] = [\tau]^\dagger [A_{TR/BR}(E)f_{TR/BR}(E,T)][\tau] \quad$ (9-a)

$n_{TR/BR}(E)$, $A_{TR/BR}(E)$ and $f_{TR/BR}(E,T)$ are known to be the filled correlation function, spectral function and fermi-dirac distribution of the reservoirs respectively. The additional factor of '2' in the numerator of equ.8 is for incorporating Kramer's degeneracy.

While self-energy due to scattering comes out to be [28,29,37]:

$$\left[\Sigma^{In}_{Scat}(E)\right] = (N_\omega + 1)[\tilde{\tau}][G^n(E + \hbar\omega_\alpha)][\tilde{\tau}]^\dagger + N_\omega[\tilde{\tau}]^\dagger[G^n(E - \hbar\omega_\alpha)][\tilde{\tau}] \quad (9\text{-b})$$

$N_\omega$ : Bose-Einstein distribution for phonons.

The modification in the interaction term ($\tau \to \bar{\tilde{\tau}}$) approaching from the reservoirs is incorporated since the local band structure at the contact domain of the reservoir is not identical to its remaining sub-part and is influenced by the interaction with neighboring domain of the active device, which leads to the modification of self-energy in equ.(9-a)[38],

$$\bar{\Sigma}^{In}_{TR/BR}(E;l,l') = b_{TR/BR}\Sigma^{In}_{TR/BR}(E;l,l') \quad (l,l' \in 1,N) \quad (9\text{-c})$$

where, $b_{TR/BR} = (1 - 2g_{TR/BR}(\beta_{TR/BR} - \bar{\beta}))$ can be termed as the enhanced coupling strength. $g_{TR/BR}$ represents the isolated green's functions of top and bottom reservoirs respectively.

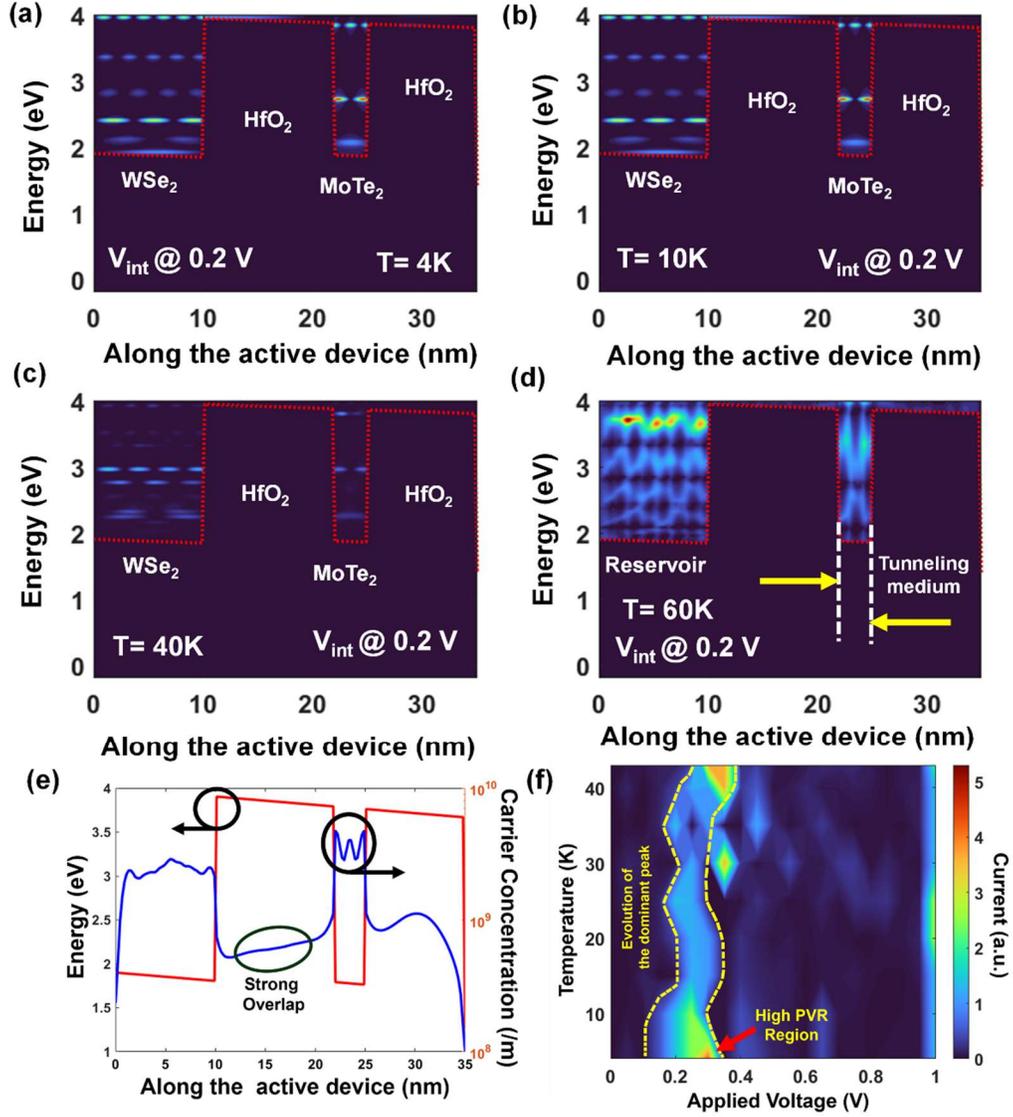

**Figure.4** Local Density of states at (a-d) 4K, 10K, 40K and 60K respectively at applied voltage 0.2V, (e) carrier concentration along the active device, and (f) current plot with applied voltage with varying temperature 4K to 45 K.

The interaction term at the contact point ($\bar{\beta}$) for the RTD structure of smaller dimension with abrupt reservoir junctions can be obtained, by employing the effective mass approximation, $\bar{\beta} = \dfrac{\beta_{AD} + \beta_{TR/BR}}{2}$ to be by using finite difference method (FDM) to construct the Hamiltonian.

The current component at m$^{th}$ lead (m=1 → Bottom Reservoir, 2→ Top Reservoir) can be estimated to be:

$$I_m(E) = \left(\frac{e}{2\pi\hbar}\right) Tr\left\{\left[\overline{\overline{\Sigma}}_m^{In}(E)\right]\left[A_D(E)\right] - \left[\overline{\overline{\Gamma}}_m^{In}(E)\right]\left[G_D^n(E)\right]\right\} \quad (10)$$

Hence, the measured resultant current comes out to be,

$$I_{INT} = \int dE \left(I_1(E) - I_2(E)\right) \quad (11)$$

From LDOS contour plots (**Figure.4. (a-d)**), it is possible to conclude that the distinguishability of the energy eigen states reduces as temperature increases, due to increment in intra-sub band transitions (IST) by means of phonon for having higher throughput deformation potential compared to other TMDC familiar materials. Such phenomena (IST) give the insinuation via shifting the dominant resonant current peak (**Figure.4. (f)**) towards higher/lower voltage. However, PVR (theoretically estimated) was found to be nearly consistent with the experimental value, still little variation in PVR & NDR (experimental) may be found because of variation in tunneling medium layer numbers, which is completely an experimental challenge to overcome. Moreover, it is noteworthy to mention that the choice of HfO$_2$ as the tunneling barriers has been done due to its higher tunneling probability for lesser electron effective mass [39,40] (even if it is 10 nm thick) which causes the strong overlap of wave functions approaching from the reservoirs to the tunneling medium, thereby signifies that an ample amount of carriers inflow from WSe$_2$ to MoTe$_2$ (identified by, transmission parameter, $\Gamma_1$ [41]) and MoTe$_2$ to the metal (similarly, $\Gamma_2$) upon application of external voltage bias.

## V. Conclusion

In this letter, we have optimized the growth of 2H-MoTe$_2$ (semiconducting-phase) over HfO$_2$/WSe$_2$ to fabricate a 2D material based resonant tunneling (RT) structure. This work successfully achieves the transverse RT conductance spectroscopy in MoTe$_2$ with PVR-4 in the low voltage range for the very first time, to be utilized for quantum sensing & metrology applications. This project can be extendable for deeper insight of such phase coherence between multiple quantum wells (2D sheets) as well as resonant tunnel assisted HEMT applications since MoTe$_2$ over other TMDC materials can be the potential candidate for high mobility/ballistic applications via controlled growth of mixed 1T' and 2H phases that overall increases the current value.

## Acknowledgements

The authors thank SPR/2023/000314, SUPRA-ANRF and National Quantum Mission, DST for financial support and infrastructural advancement. Authors show their gratitude to the DRDO Industry Academia Centres of Excellence (DIA-CoEs) for infrastructural development. The authors express gratitude to the Ministry of Education (MoE), Govt. of India, with grant no. MoE-STARS/STARS-2/2023-0621 for financial support. The authors also extend their gratitude to the Ministry of Human Resource Development (MHRD), India, for funding this work through the 'Grand Challenge Project on MBE Growth of 2D Materials' (MI01800G). Authors are grateful to Central Research Facility (CRF), Sophisticated Analytical & Technical Help Institutes (SATHI) and Nanoscale Research Facility (NRF), IIT Delhi for infrastructural support. Authors acknowledge Dr. Atul Kumar Singh, CRF, IIT Delhi and Mr. Suprovat Ghosh, CARE, IIT Delhi, for fruitful discussions.

## Authors Contributions

**A.M.:** Conceptualization, Theoretical Studies, Materials' Growth & Characterization, Device Fabrication, Electrical Measurements and Writing Original Draft, Editing & Review; **K.S.:** Materials' Growth & Characterization, Device Fabrication and Electrical Measurements and Original Draft Editing & Review; **K.B. and S.K.:** Materials' Growth & Characterization and Original Draft Review; **S.D. & R.S.:** Funding and Supervision.